\documentclass[prl,floatfix,superscriptaddress,nofootinbib,twocolumn]{revtex4}
\usepackage{dcolumn,amsmath}
\usepackage{graphicx}
\usepackage{bm}
\usepackage{amssymb}
\usepackage{hyperref}
\usepackage{multirow}

 \newcommand{\Table}[1]{Table~\ref{#1}}
 \newcommand{\Fig}[1]{Fig.~(\ref{#1})}
 \newcommand{\Figs}[1]{Figs.~(\ref{#1})}
 \newcommand{\Eq}[1]{Eq.~(\ref{#1})}

 \newcommand{\etal}{{\it et~al.}\ }
 \newcommand{\abinitio}{{\it ab~initio} }
 
 \newcommand{\Eeff}{$E_{\rm eff}$}
 \newcommand{\eEDM}{{{\it e}EDM}}
 \newcommand{\Apar}{$A_{\parallel}$}
 \newcommand{\Aperp}{$A_{\perp}$}

\usepackage{xcolor}
  \newcommand{\comtav}[1]{\pdfcomment[color=red,author="TAV"]{TAV:#1}}
  \newcommand{\comyul}[1]{\pdfcomment[color=green,author="YuL"]{YuL:#1}}
  
\usepackage[final]{pdfcomment}

\begin{document}
\title{Concept of effective states of atoms in compounds to describe properties determined by the densities of valence electrons in atomic cores}
\author{Anatoly V. Titov} \email{anatoly.titov@gmail.com}
                          \homepage{http://www.qchem.pnpi.spb.ru}
\author{Yuriy V.\ Lomachuk}\email{jeral2007@gmail.com}
\author{Leonid V.\ Skripnikov}\email{leonidos239@gmail.com}

\affiliation{B.P.\ Konstantinov Petersburg Nuclear Physics Institute, Gatchina, Leningrad district 188300, Russia}
\affiliation{Department of Physics, Saint Petersburg State University, Saint Petersburg, Petrodvoretz 198504, Russia}

\date{\today}

\begin{abstract}
        \comyul{novelty}
We propose an approach for describing the effective electronic states of ``atoms in compounds'' to study the properties of molecules and condensed matter which are circumscribed by the operators heavily concentrated in atomic cores. Among the properties are hyperfine structure, space parity (P) and time reversal invariance (T) nonconservation effects, chemical shifts of x-ray emission lines (XES), M\"{o}ssbauer effect, etc. An advantage of the approach is that a good quantitative agreement of predicted and experimental data can be attained even for such difficult cases as XES chemical shifts providing correct quantum-mechanical interpretation of the experimental data. From the computational point of view the method can be quite efficient being implemented in the framework of the relativistic pseudopotential theory [A. V. Titov and N. S. Mosyagin Int.J.\ Quantum Chem.\ {\bf 71}, 359 (1999)] and procedures of recovering the wave functions in heavy-atom cores [A. V. Titov, N. S. Mosyagin, A. N. Petrov,
and T. A. Isaev, ibid.\ {\bf 104}, 223 (2005)] after a molecular, cluster or periodic structure calculation performed on the basis of pseudoorbitals smoothed near the nuclei within the pseudopotential approximation.
We report results of our studies of a number of atomic and molecular systems to demonstrate the capabilities of the approach.
%
\end{abstract}

\maketitle


\section{Introduction}
\label{intro}


One of the most popular ideas in quantum theory of electronic structure of molecular and condensed matter systems is the concept of {\it atoms in molecules} (AiM). Its use allows one to understand some chemical-physical properties of a whole system by analyzing the characteristic properties of constituent atoms. Though there is a well-known formulation of ``quantum theory of atoms in molecules'' developed by Bader \cite{Bader:94, Bader:98}, unambiguous and commonly accepted definitions of such terms as ``partial atomic charges'' or ``state of an atom in a molecule'' do not exist. Using different theoretical backgrounds and pursuing certain goals one derives different results, which can be useful for some applications and not so useful for the others (e.g., see the discussion about partial atomic charges on p.~309 of Ref.\cite{Cramer:04}).

%


Several basic concepts and quantitative tools are widely exploited in the literature on the subject.
{\it Basic concepts}
such as the oxidation state (number), valence, formal charges, etc.\ are commonly used to characterize the charge states of atoms in molecules to get a preliminary idea about the chemical structure of a compound of interest.
 \comtav{Referee: For example, when it is said that quantities like oxidation state, valence are not correctly defined from quantum mechanical point of view, what does it mean? <<<}
%
%
     However, there are no well-defined representations of these concepts by observable quantities which would be commonly accepted by the physical-chemical community.
  This means that the following applies for ``intuitively useful'' quantities such as oxidation state, valence etc.\ 
{\bf (i)} They cannot be presented as expectation values of some
  unambiguous quantum mechanical
operators; %
{\bf (ii)} they cannot be uniquely determined
from experiments (even nominally, see \Table{table:tabl2} for different charge states of Pb causing the same chemical shifts);%
{\bf (iii)} they can be well-defined theoretically and/or experimentally but not very helpful from practical point of view to be used for
analysis of vital electronic properties of a chemical system (see next paragraph and ``Class III charges'' in \cite{Cramer:04}).

\comtav{>>>End of change} 
{\it Quantitative approaches}
 to describe the effective states of atoms in compounds are mainly based on using
 Hartree-Fock, natural or localized orbitals and one-electron density matrices; alternatively, they
are originated on analysis of the total electronic density $\rho(\vec{r})$ of a chemical system, utilizing spatial criteria or reproducing some experimental data within simple theoretical models (e.g., see Ref.~\cite{Cramer:04} and text below).

Each of the known definitions has not only advantages but drawbacks which can seriously weaken the former in specific applications. In particular, the methods based on the
Mulliken and L\"{o}wdin population analyses \cite{Lowdin:70} strongly depend on the basis set used and are not so useful for large basis sets. The more elaborate concept of natural atomic orbitals (NAOs) \cite{Reed:88, Weinhold:05} overcomes the problem but the valence NAOs of an atom can notably differ in various chemical compounds and they are generally not localized on an atom, thus complicating the comparison of effective states of the atom in different molecular environments in terms of occupation numbers, etc.
\comtav{??? corrected <<<}
The methods utilizing the electronic densities directly, including those by Hirshfeld \cite{Hirshfeld:77}, Bader (see Ref.~\cite{Bader:94} and next paragraph), Voronoi, etc.\ (see Refs.~\cite{Cramer:04, Ramos-Cordoba:13}), suggest partitioning of a chemical compound
  by using some spatial criteria in a manner that
    is not an intrinsic quantum mechanical
    hallmark
even if it is compatible with quantum mechanics description of the compound.
\comtav{>>> End}
%
The auxiliary concepts introduced by such a way are logical when determining the charge states of a given atom for particular cases only.

The most sensible of the methods for determining the effective states of atoms (and not just their partial charges) in chemical compounds described in the literature are anyway based on evaluation of some kind of overlapping the electronic wave function of a chemically bonded atom with those of a free atom and its ions. In other words, a mapping of the effective state of an atom in a molecular environment on states of the free atom takes place.
%
One of the natural approaches of this group described in the literature is recently formulated in Ref. \cite{Ramos-Cordoba:13}. The ``charge states'' of atoms in a molecule are determined there 
  when assigning to an atom the parts of molecular orbitals truncated with using the Bader zero-flux surfaces.
  \comtav{Referee: Similarly, when the authors talk about zero flux
condition of Bader, it would be good to write in a couple of lines
what it implies. <<<}
 These surfaces are defined by the equation
\begin{equation}
   \nabla\rho(\vec{r}) \cdot \vec{n}(\vec{r}) = 0,
\end{equation}
for every point $\vec{r}$ on the interatomic boundary surface, with the unit vector $\vec{n}(\vec{r})$ normal to the surface.
Thus, the atoms constituting the chemical system can be separated by the spatial criterion using total electronic density only.
\comtav{>>> End of change}
The overlap integrals of molecular orbitals with atomic functions are calculated not over the whole space, but only in the areas bounded by zero-flux surfaces. These surfaces define the boundaries of atoms in a molecule within the Bader analysis and the atomic expansion coefficients of the molecular orbitals thus obtained are, obviously, consistent with the atomic Bader charges.
However, as is discussed above, such a definition of partial atomic charges and spatial partitioning into atomic regions
%
is not completely consistent with the quantum-mechanical description of electronic structures. In particular, the relativistic atomic orbitals (spinors) with different $l$, $j$ components for valence electrons usually have notably different spatial localizations and the atomic regions cannot be separated by any surfaces unambiguously.


~



To calculate the properties of molecules described by quantum mechanical operators heavily concentrated in atomic cores but sensitive to variation of densities of valence electrons, combined (two-step) approaches have been developed \cite{Titov:96, Titov:99, Titov:05a, Skripnikov:13b} and applied to study hyperfine structure (HFS), space parity (P), and time-reversal invariance (T) nonconservation (PNC) effects.
\comtav{Table~I: <<<}     
     Earlier calculations of these properties, which are in the context of the subject of this paper are given in \Table{table:tabl1} (see also discussion in the next section).
\comtav{>>> End}
 
In Ref.~\cite{Lomachuk:13} a method of evaluating chemical shifts of x-ray emission spectra (XES) for compounds containing period four and heavier elements, that is consistent with the two step approach is also proposed.
 
On the basis of these developments
one can introduce a method for determining the effective state of a given atom
  in a chemical compound (substance).

This method is originated from the relativistic pseudopotential theory \cite{Titov:99, Petrov:04b, Mosyagin:06amin} and one-center restoration approaches \cite{Titov:05a, Titov:06amin} to recover proper electronic structure in heavy-atom cores after the relativistic pseudopotential simulation of a chemical substance. The present research can also be considered as a generalization of our computational models utilized to study the HFS and PNC effects as well as XES chemical shifts in molecules and solids.





\section{Motivation}
\label{sec1}



The observable properties and effective Hamiltonian parameters of our interest here include those that can be measured and those that can only be calculated and, thus, usually need to be checked by some appropriate way. The group of measurable properties comprise magnetic dipole and electric quadrupole HFS constants, XES chemical shifts, isomeric (chemical) shifts of M\"{o}ssbauer spectra, the volume effect of isotopic shifts. The unmeasurable effective Hamiltonian parameters cover those required to study T- and P-nonconservation effects in nuclei, atoms, molecules and solids: effective electric field \Eeff\ on (unpaired) electrons required for the electron electric dipole moment (\eEDM) search; electronic density gradients on nuclei for Schiff moment; electromagnetic field on nuclei for anapole moment etc. Some examples of these studies are discussed in the following three paragraphs.


In Refs.~\cite{Dmitriev:92, Mosyagin:98, Isaev:04, Petrov:05a, Isaev:05a} the calculated HFS constants in various compounds are compared to the corresponding experimental data \cite{Knight:81, Mawhorter:11, Hunter:02, Kawall:05, Chanda:95} to estimate the errors in the effective electric field calculations of HgF, PbF, YbF, HI$^+$, and PbO, see \Table{table:tabl1}.

The XES lines correspond to electronic transitions between the shells localized in atomic cores. Comparing the experimental and theoretical XES data one can analyze the electronic structure (effective state) of an atom in a compound (see below). The studies intended to extract the information about the electronic structure in atomic cores in different chemical compounds from the XES data are performed, in particular, in Refs.~\cite{Raj:98, Raj:02, Pawlowski:02}. The authors estimate the $3d$ shell occupancies in various metals by comparing the measured $K\beta$ to $K\alpha$ x-ray intensity ratios to the results of atomic multiconfiguration Dirac--Fock computations.  In Refs.~\cite{Sumbaev:76, Sovestnov:09} the electronic structures of various compounds were studied by comparing the experimental data on the x-ray chemical shifts to the results of corresponding atomic calculations. 

It can be shown \cite{Dickson:05} that the chemical shifts of M\"{o}ssbauer spectra (isomer shifts) in various compounds are proportional to the differences of the total electronic densities on a given atomic nucleus. The measured isomer shift values as functions of the oxidation states (ligands) of the Fe, K, Ir, As,
\comtav{Ph <<<}
   and other
\comtav{>>> End}
atoms in different compounds are given in Ref.~\cite{Dickson:05} together with the calculated electronic densities on the atomic nuclei as functions of electronic configurations for the neutral and ionized Fe and Sn atoms.

The discused properties and parameters characterize the processes occurring in atomic cores or, for stationary states, they are mean values of the operators heavily concentrated on nuclei or in atomic cores. By other words, the properties and parameters of our interest
%
 strongly depend on
the electronic configuration ({\it effective state}) of a given {\it atom in a compound} (AiC)%
\footnote{We will further use this terminology and acronym AiC to distinguish them from the widely used terms ``atoms in molecules'' and AiM.}
rather than on the {\it chemical bonds} between atoms.
%
We call such characteristics the {\it AiC characteristics} or {\it AiC properties and parameters} (together with the ``core characteristics'', ``core properties'', and ``core parameters'' as in our earlier papers) assuming that the processes and quantum mechanical operators considered are spatially localized near nuclei despite the fact that not core but valence electrons usually give a key contribution to the given properties and parameters.

The ultimate aims pursued in our consideration are:
(i)~to formulate a robust model for description of the effective states of atoms in compounds;\\
%
(ii)~to attain close and unambiguous connection between the quantities which can be measured (chemical shifts of x-ray emission lines etc.) and theoretical models for their evaluation;\\
%
%
(iii)~to give insight into the quality of \abinitio calculations or semiempirical estimates of the AiC characteristics which cannot be (or are not yet) measured;\\
%
(iv)~to provide a unified tool for indirect (or ``independent'') accuracy check of the evaluated AiC characteristics;\\
(v)~to give a theoretical background for development of advanced (combined) computational schemes, which would be optimal (in the ratio quality to price) for their study;\\
(vi)~to make calculations more feasible (easier, faster, and more reliable) for computationally difficult cases (e.g., for complicated molecules and condensed matter structures containing heavy $d$ and $f$ elements).

It is well known that not all the well observable properties can be used for testing the calculated AiC characteristics but only those that have comparable sensitivity to variation of the electronic densities (or, generally, density matrix) in the vicinity of a nucleus due to electronic structure reorganization from one compound to the other, perturbations or electronic excitations in the valence region, etc. On the other hand, some properties that can serve as a good check for a given AiC characteristic in one kind of compounds are not suitable for the other ones. As an example, for such molecules-radicals as BaF, YbF and HgF \cite{Titov:06amin} with $sp-$hybridized state of unpaired (valence) electron, a good semiempirical estimate for \Eeff\ can be written as \Eeff${\sim}\sqrt{A*A_d}$, where $A=(A_{\parallel}{+}2A_{\perp})/3, A_d=(A_{\parallel}-A_{\perp})/3$, $A_{\parallel}$ and $A_{\perp}$ are magnetic dipole hyperfine structure parameters
\cite{Kozlov:95}. However, this formula is not so useful for the systems with $d$ and $f$ unpaired electrons only \cite{Skripnikov:11a}. So, a systematic analysis of applicability of some (measurable) properties to test the other (unmeasurable) ones is required. 

The results of our earlier calculated values and experimental data for 
$A_{\parallel}$, $A_{\perp}$ and electric quadrupole hyperfine structure constant $eQq_0$
in HgF, PbF, YbF, HI$^+$, and PbO molecules are presented in \Table{table:tabl1}.%
\comtav{Referee: I also found that presenting Table-I did not really
serve any purpose. Modified: <<<}
   As one can see, the calculated HFS values are within $10-15$\% agreement with the experimental data. We should note that the calculations given in \Table{table:tabl1} are performed with using the approximations that are compatible with the theory given in the next section, whereas accuracy of modern, more sophisticated calculations is notably better (e.g., see recent results on ThO in Ref.~\cite{Skripnikov:14b}).
\comtav{>>> End}     

\begin{table*}[t!]
\begin{minipage}{\textwidth}
\caption{\label{table:tabl1}Hyperfine structure constants.}
\begin{ruledtabular}
\begin{tabular}{clcccc}
& & \Apar, MHz & \Aperp, MHz & $eQq_0$, MHz &  Relative error\\
\hline
\\
\multirow{2}{*}{$^\mathrm{199}$HgF}& theory, Dmitriev \etal (1992) \cite{Dmitriev:92} &24150&23310&  &      15\%\\
                   &expt., Knight \etal (1981) \cite{Knight:81} &22621&21880&     &     \\
                   \\
\multirow{2}{*}{$^\mathrm{207}$PbF} &theory, Dmitriev \etal (1992) \cite{Dmitriev:92}&10990& -8990&&15\%\\
              &expt., Mawhorter \etal (2011)\cite{Mawhorter:11}&10147& -7264& &\\
              \\
\multirow{2}{*}{$^\mathrm{171}$YbF} &  theory, Mosyagin \etal (1998)\cite{Mosyagin:98}&  8000&  7763&&15\%\\
              &expt., Steimle \etal (2007)\cite{Steimle:07}&  7424&  7178&&\\
              \\
   \multirow{2}{*}{H$^{\mathrm{127}}$I$^+$} & theory, Isaev \etal (2005)\cite{Isaev:05a}&    968&&-745&10\%\\
              &expt., Chanda \etal (1995) \cite{Chanda:95}&  1021& &-712.6&\\
              \\
\multirow{2}{*}{$^\mathrm{207}$PbO$^*$, a(1)} & theory, Petrov \etal (2005) \cite{Petrov:05a,Isaev:04}&-3752&& &10\%\\
        &  expt., Hunter \etal (2002) \cite{Hunter:02}  &   -4110(30)\\
 \multirow{2}{*}{$^\mathrm{207}$PbO$^*$, B(1)} & theory, Petrov \etal (2005) \cite{Petrov:05a,Isaev:04}&4965\\
        & expt. (after theory), Kawall \etal (2005) \cite{Kawall:05}  &   5010(70) && \\
       
\end{tabular}
\end{ruledtabular}
\end{minipage}
\end{table*}

\section{Theory}
\label{sec2}

According to the above-discussed motivation we are going to formulate the AiC model for applications that satisfy the following basic criteria:
  (i) correct quantum mechanical description;
 (ii) common features of the AiC characteristics should be taken into account;
(iii) a good {\it quantitative} agreement of the AiC-theory predictions with
  experiment.

The importance of the first criterion is discussed above, first of all it is concerns the chemical shifts, for which the present status of theory cannot be considered as satisfactory.
Taking into account the common features of the AiC characteristics is discussed below; it assumes exclusion of those computational elements from the model which do not affect essentially the AiC properties and parameters, allows one to reduce the computational cost visibly and makes the calculation more transparent.
 
As to the latter criterion, for different AiC characteristics and various kinds of compounds the term ``a~good agreement'' can have different meanings. Consider as an example the XES chemical shifts for heavy elements discussed in the introduction. They can be three to six orders of magnitude smaller than the energies of $K,L$ lines and the difficulty of their evaluation is highly aggravated by the two-electron nature of energetic properties \cite{Lomachuk:13}. Therefore, calculation of such chemical shifts with a good accuracy is a very serious problem in practice. In general, however, we assume the disagreement of an AiC-model prediction with experiment should be at most on the level of ${\sim}30$\% to provide a satisfactory qualitative description of an experiment, though, a benchmark \abinitio\ calculation can provide, in principle, a better accuracy.
\comtav{Added: <<<}
     Note here that typical errors, $10{\div}15$~\%, for the HFS constants (\Table{table:tabl1}) and error estimations for PNC effects \cite{Titov:06amin}, which were earlier evaluated within simple two-step models \cite{Titov:05a}, are also compatible with this limitation.
That is important since one of main goals of the AiC theory is to provide accuracy check for those {\it unmeasurable} molecular parameters which are required to study PNC effects.
\comtav{>>> End}


The principal common feature of the AiC characteristics given above is that the direct contributions from the spatially valence region ($r{>}R_c$, the choice of the core radius $R_c$ is discussed in the next paragraph) to their values are relatively small, though, these are the valence electrons which determine the effective state of a free or bounded atom and mainly control the AiC characteristics taking into account the inactivity of core shells. The valence states contribute directly to these properties by only their small parts with the electronic density share ${<}10$\% localized in the atomic core ($r{<}R_c$) and not by the valence and outer regions ($r{>}R_c$) with the share ${>}90$\%
\comtav{Referee: it is not clear how the inter-atomic region enters into the
calculations. For example, if the core wavefucntions (spinors) of
atoms are affected by the outside regions so that the density at the
nucleus change, the cause of this change is non-transparent.}
     \footnote{In calculations of heavy-atom compounds the core shells of the heavy atom(s) are usually treated as frozen. However, one can partially account for a relaxation of the corresponding core shells of an atom in different compounds by performing calculations of the isolated atom after determining the state of the atom in the compound. We use such approximation for estimating XES chemical shifts values in various lead compounds below in the paper (see~\Table{table:tabl3} for details).}.
\comtav{<<<End of change}

In turn, the native atomic potential from the nucleus and core electrons is ``hard'' for the valence electrons in its own atomic core, i.e., it is much higher by the amplitude than the potentials of other atoms or external sources. It is also much more by amplitude than the energies of the one-electron {\it valence and low-lying virtual} (W) states. Note that the valence and low-lying virtual states may change places with each other in different environments of the atom, processes etc., so one should treat them on equal footing when studying the effective states of an atom in different compounds.
We take account of relativistic effects and, thus, use spherical spinors
($\Omega_{ljm}(\vec{r}/r)$ or $|ljm\rangle$,
where $l,j$~are the orbital and total angular momentum quantum numbers, $m$ is the projection of $j$) for the spin-angular part of W states.
Neglecting the outer potentials and energies of the W states in the atomic core, 
the property of proportionality or homogeneous scaling of the W spinors takes place in the core ({\it W proportionality} below) as is shown in \Figs{fig1} and (\ref{fig2}) \cite{Petrashen:56, Flambaum:90, Titov:99, Titov:02Dis}.
\comtav{Referee: Furthermore, it is stated in the context of Fig. 2 that the valence
and virtual spinors are proportional to each other but it is not
explained what its implications are. <<<}
  This property allows one to introduce the {\it W reduced density matrix} (given below)
     that leads to a more intuitive
       (basis-set independent and ``minimally-sufficient'')
formulation of the density matrix concept to evaluate the AiC characteristics (that is discussed in detail below).
\comtav{>>> End}
\begin{figure}[h!]
 \caption{Large components of the $6p_{1/2}$, $7p_{1/2}$ spinors of Pb for the $5s^2\,5p^6\,5d^{10}\,6s^26p^2$ configuration. The  large components of $6p_{1/2}$ and scaled $7p_{1/2}$ spinors in the core region are given in subfigure, where the scaling factor is chosen in such a way that the amplitudes of large components of these spinors are equal at $R_c=0.5$~a.u.}
  \centering
    \includegraphics[width=\columnwidth]{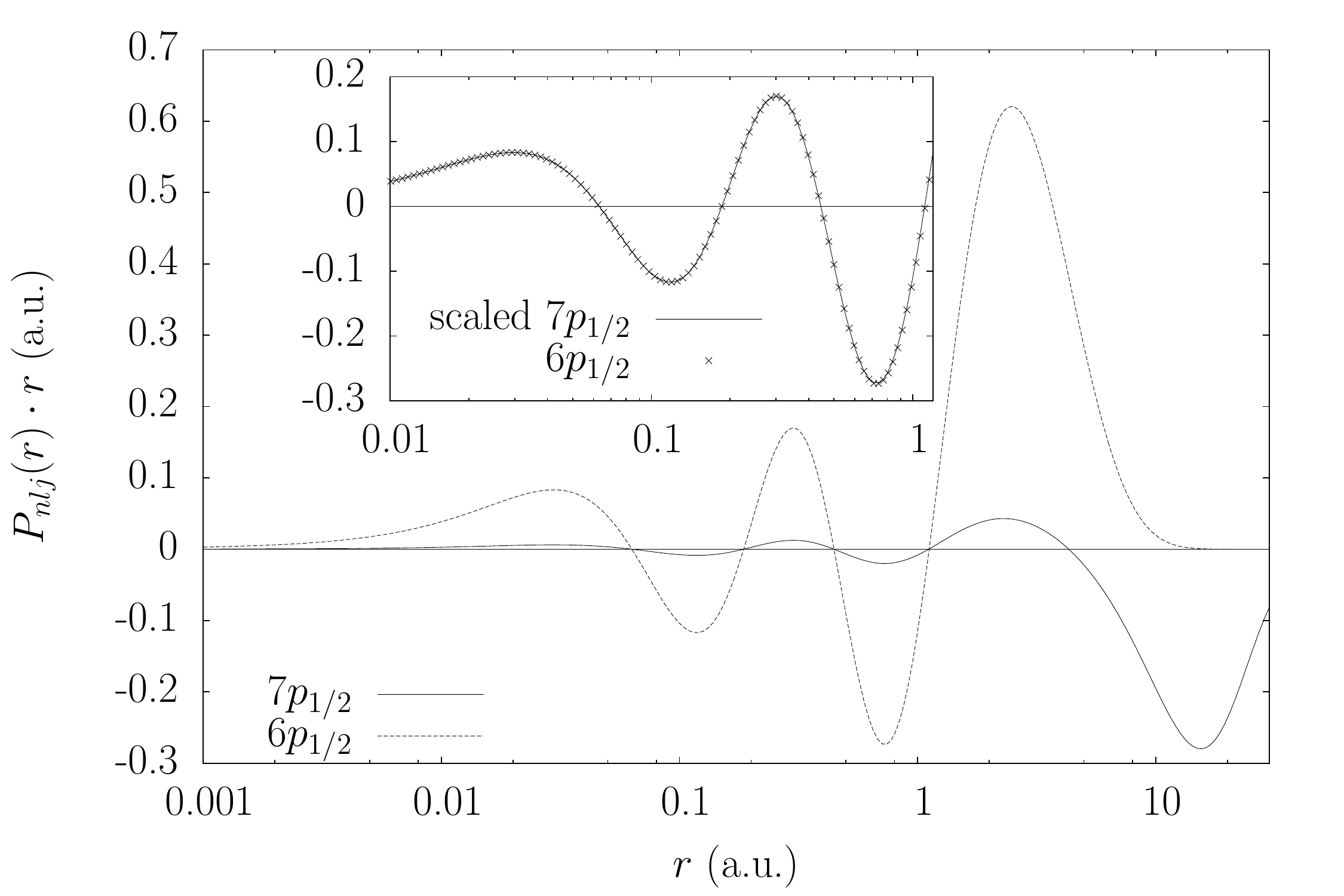}
  \label{fig1}
\end{figure}
%
\begin{figure}[h!]
  \caption{Large components of the $6p_{1/2}$ spinors of Pb for $[5s^2\,5p^6\,5d^{10}]6s^2\,6p^2$, $[\dots]\,6s^{1.11}\,6p^2$, and $[\dots]6s^{0.53}\,6p^{0.57}$ configurations. The first one corresponds to the ground state and the next two are roughly (according to Ref.~\cite{Kaupp:93}) equivalent to the states of Pb in the PbH$_4$ and PbF$_4$ molecules, correspondingly.}
  \centering
    \includegraphics[width=\columnwidth]{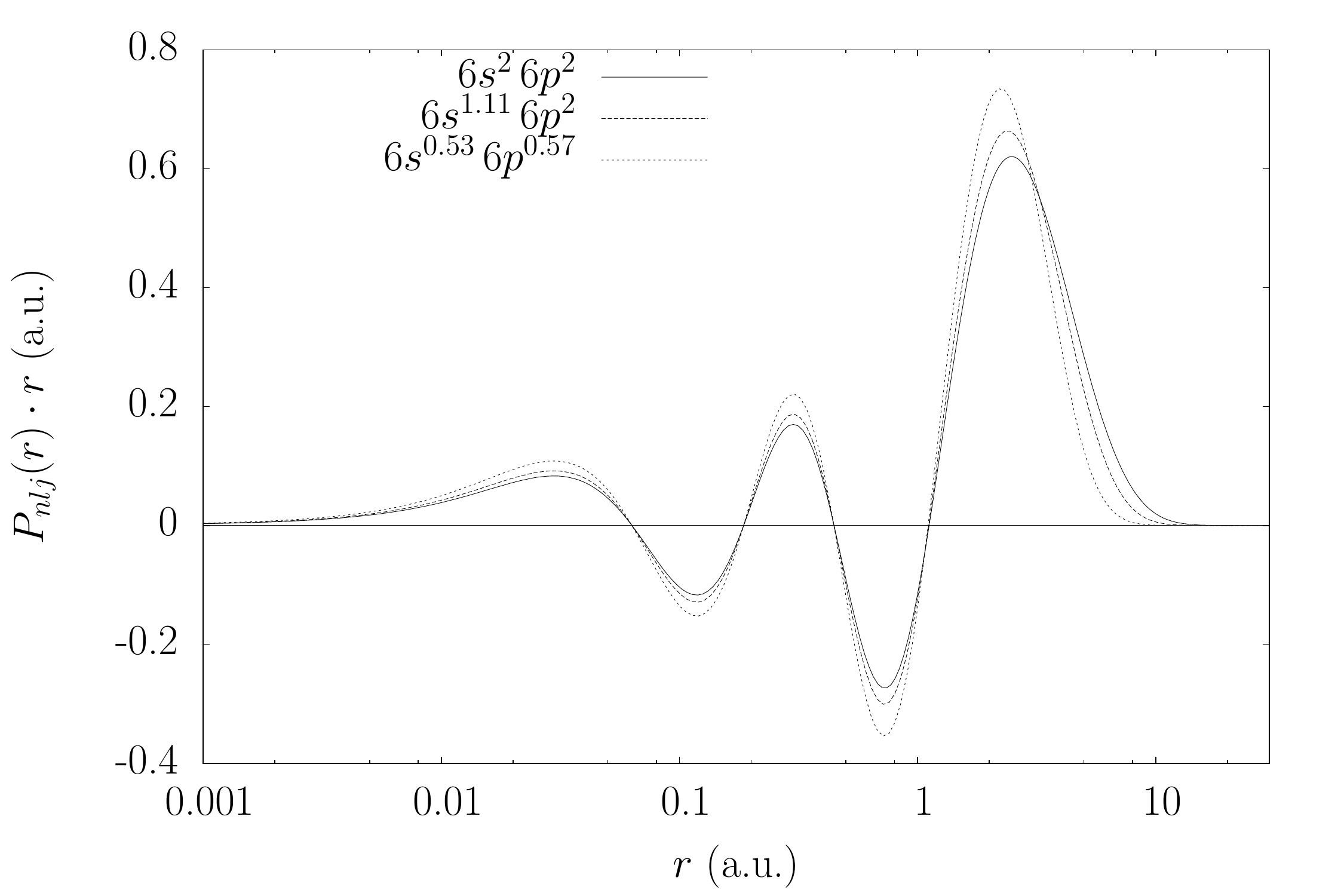}
  \label{fig2}
\end{figure}
One can see from \Fig{fig1} that the valence and virtual spinors are proportional each other in the atomic core. \Fig{fig2} illustrates the proportionality of the valence spinor generated for different states of the atom.


The theoretical backgrounds of the AiC approach can be formulated as follows:
 ({\bf i}) the core radius $R_c$ is chosen for a given atom in a compound by such a way that the contributions to a characteristic of interest outside the sphere with this radius ($r{>}R_c$) can be neglected or small enough and valence spinors with same angular quantum numbers are proportional to each other
\comtav{Referee: Similarly, it is not apparent how $R_c$ is fixed, although the authors
do say in their conclusion that $R_c$ does not affect the results. <<<}%
\footnote{%
From a formal point of view, the smaller $R_c$, the more unambiguously $W$ proportionality can be defined. In practice, however, one needs to consider computational errors
   at small $R_c$.
%
Besides, the larger radius of an AIC operator, the partial waves
with higher values of the angular quantum numbers $j$ and $l$
one should treat, in general, to attain appropriate accuracy for the characteristic of interest.
In applications such calculations are performed using (relativistic) pseudopotentials with the pseudoorbitals (pseudospinors) smoothed in some core region within a matching radius (detailed studies on the subject are given in \cite{Titov:99}). Within the hard-core (accurate) pseudopotential formulations the matching radii should be as small as possible. In practice, they are close to the last (by amplitude) maxima, $R^V_m$, of valence orbitals (thus the ``large-core'' pseudopotentials are generated) or, for better accuracy, to the last maxima, $R^{OC}_m$, of outermost core orbitals (for the ``small-core'' pseudopotentials). So,
   for the all-electron four-component case
one can write
$R_c\le R^V_m$, 
   whereas, for the case of the pseudospinors smoothed in the atomic core, one should first restore proper four-component behavior of spinors in the core or use $R_c$ at least not less than the largest matching radius
\cite{Titov:05a}.};
\comtav{>>> End}
 ({\bf ii})
the W proportionality in the core is applied to generate unique relativistic
(four-component) reference spinors, $\eta_{ljm}(r)$ for $r{<}R_c$, which are not smoothed in the core. Being normalized within the sphere with radius $R_c$:
\begin{equation}
   \int\limits_{r<R_c} r^2 dr |\eta_{ljm}(r)|^2 = 1\ , 
 \label{eq:orthog}
\end{equation}
these functions become universal and almost independent from the state (states) of an atom and its ions for which they are constructed; here we assume the ground or low-lying states of the atom with excited or ionized valence electrons only. Consider further the set $\{\eta_{lj}(r)\}_{lj}$ such generated as the ``AiC basis'' for a given atom to describe its effective state in a chemical substance.

Designate the four-component W spinors on a given atom as
\begin{equation}
 \varphi_{n_wljm}(r) =
  \left(\begin{array}{l}
    \varphi^{L}_{n_wljm}(r)\\
    \varphi^{S}_{n_wljm}(r)
  \end{array}\right) ,
 \nonumber
\end{equation}
in accord to the conventional representation of atomic Dirac-Fock spinors, $\varphi_{n_xljm}(r)$, by means of large (L) and small (S) components.
The index $n_x$ can naturally be the principal quantum number not only for core ($n_c$) and valence shells but for the virtual ones when a finite and localized basis set or/and an external spherical wall-type potential on the atom is used to generate atomic spinors with $n_x$ that grows monotonically with the energies of $\varphi_{n_xljm}(r)$ states. Now the W states are those that match the appropriate choice of $n_w: n_c{<}n_w{<}n_r$, where the index
$n_r$ corresponds to the highly excited atomic states with energies comparable or more than the amplitude of the atomic potential within the sphere with radius $R_c$.

Thus, the W spinors in the core region with $r{<}R_c$ can be written as 
\begin{equation}
    \varphi^{<}_{n_wljm} (r) = k_{n_wljm} \eta^{\vphantom{W}}_{lj}(r)\, ,\ r{<}R_c\ ,
 \label{eq:valorb}
\end{equation}
since the radial parts of W spinors with the same $l,j$ and different $m$ are also proportional each other in the core. 


With this background the following reducing of a one-electron density matrix that describe the atomic, molecular and condensed matter structures can be performed:

(1) the valence and low-lying virtual molecular/crystal orbitals or spinors are reexpanded on a basis set of atomic (one-center) W spinors within the sphere with radius $R_c$:
\begin{equation}
\begin{split}
 \psi^{W<}_i(\vec{r}) \approx \sum\limits_{n_wljm} c^i_{n_wljm}
  \left(\begin{array}{l}
    \varphi^{L}_{n_wljm}(r)\Omega_{l,jm}(\vec{r}/r)\\
    \varphi^{S}_{n_wljm}(r)\Omega_{2j{-}l,jm}(\vec{r}/r)
  \end{array}\right) , \\
|\vec{r}|{<}R_c\ ,
\end{split}
 \nonumber
\end{equation}
where $\Omega_{l,jm}$ and $\Omega_{2j{-}l,jm}$ are the conventional spin-angular factors for large and small components of an atomic bispinor.

(2) the one-electron density matrix (DM) of these spinors,
${\bm\rho}^W[\psi^{W<}_i]\equiv(\rho^{<}_{n_wljm,n_w'l'j'm'})$,
is calculated and reexpanded on the chosen atom using a one-center basis set. The ${\bm\rho}^W$ matrix can be obtained from the one-electron density matrix,
   ${\bm\rho}$,
that describes all electrons of the system as follows:
\begin{equation}
  \begin{split}
        \bm{\rho}^{W}= P_\mathrm{W}\bm{\rho}P_\mathrm{W} \equiv\\
    (1-P_\mathrm{C}-P_\mathrm{R})\bm{\rho}(1-P_\mathrm{C}-P_\mathrm{R})\ ,
  \end{split}
 \label{eq:rho_w_begin}
\end{equation}
where $P_\mathrm{W}$ is the projector on the W states only, $P_\mathrm{C}$ and $P_\mathrm{R}$ are the projectors on the core states of a selected atom and the states having negligible densities in the core of this atom, correspondingly.
Note that none of both diagonal and off-diagonal submatrices between the $P_C, P_W$ and $P_R$ projectors are zero in general
when correlation, relaxation and mixing of different harmonics take place for an atom in a chemical substance. However, the core shells (particularly, for the small-core cases) can usually be treated as frozen atomic spinors with a very high accuracy, the DM will be diagonal on the core states and the corresponding submatrix can be safely removed from consideration even for XES chemical shifts \cite{Lomachuk:13}.
Moreover, the space of R states describing mainly the core relaxation and correlation effects do not contribute to the DM in these cases, they can be neglected and the \Eq{eq:rho_w_begin} can be rewritten as
\begin{equation}
  \begin{split}
        \bm{\rho}^{W} \approx\ (1-P_\mathrm{C})\bm{\rho}(1-P_\mathrm{C})\ .
  \end{split}
 \label{eq:rho_w_mid}
\end{equation}
 The only a question in practice is to partition the one-electron states on the C, W and R subspaces correctly to minimize computational efforts for an accuracy of the interest.
%
Taking account of \Eq{eq:valorb} the $\bm{\rho}^{W}$ matrix can be transformed to a new one:
\begin{equation}
  \Delta_{ljm,l'j'm'} = \sum\limits_{n_w n_w'}k_{n_wljm}k_{n_w'l'j'm'}\rho_{n_wljm,n_w'l'j'm'}.
 \label{eq:red_dens}
\end{equation}
   The new matrix is already reduced (summed up) on the principal quantum numbers $n$ for only W spinors which are taken into account, whereas the core and high-energy virtual states are excluded, so, call this matrix as the {\it W reduced DM}.


Because of the W proportionality, a physical meaning for the AiC characteristics has only the W reduced DM since it is generally impossible to distinguish the distribution of electrons by W spinors with fixed $l,j,m$ numbers in atomic cores by any available data on AiC properties. Moreover, partitioning the W space on individual states has a meaning only for free or weakly bound atoms, it is almost meaningless for a chemically bound atoms and, particularly, for condensed matter structures.

Let us consider the diagonal terms $\Delta_{ljm,ljm}$ of the W reduced DM. Multiplying these terms by the charge of the corresponding reference states $\left\{\eta_{ljm}\right\}$ within the sphere of radius $R_c$, one obtains the $\left\{q^{W}_{lj}\right\}$ quantities for a given atom taking into account \Eq{eq:orthog}:
\begin{equation}
     q^W_{lj} = \sum_m \Delta_{ljm,ljm} \int\limits_{r<R_c} r^2 dr |\eta_{ljm}(r)|^2 = \sum_m\Delta_{ljm,ljm}\ ,
 \label{eq:partialwavecharges1}
\end{equation}
 which we call below the core region {\it partial wave charges}.

Alternatively, as one can easily see, the partial wave charges of an atom ``$A$'',
$q^{WA}_{lj}$, may be defined as the expectation values of the projection operators 
\begin{equation}
  P^{<,A}_{lj}=\sum_m\left|ljm\right\rangle\theta(R_c-|\vec{r}-\vec{R_A}|)\left\langle ljm\right|
 \label{eq:Abar_tailless}
\end{equation}
on the W density matrix $\bm\rho^W$:
\begin{equation}
q^{WA}_{lj} = \mathrm{Tr}\left[P^{<,A}_{lj}\bm\rho^W\right].
 \label{eq:partialwavecharges1}
\end{equation}
The Heaviside step function $\theta(R_c-|\vec{r}-\vec{R_A}|)$ is equal to unity in the core of atom ``A'' and zero outside:
\[
\theta(R_c - |\vec{r}{-}\vec{R_A}|) = \left\{\begin{array}{l}
1,\  |\vec{r}{-}\vec{R_A}|<R_c\\
0,\  \mbox{otherwise}
\end{array}\right. ,
\]
and $\left| ljm\right\rangle$ is the spherical spinor discussed above.

The operator (\ref{eq:Abar_tailless}) can be interpreted also as a ``tailless'' semilocal model potential (pseudopotential) of 
Abarenkov-Heine type \cite{Abarenkov:65}
 that is independent on the radial parts features of atomic spinors.
 Thus, it can be easily utilized in most available quantum chemical codes to evaluate the partial wave charges on an atom. The only a limitation is in its using together with the DFT wavefunctions which do not allow one to construct DM correctly
but the electronic density only, so the accuracy in DFT calculations of partial wave charges can be, in principle, quite low.


\begin{table*}
        \begin{minipage}{\textwidth}
                \caption{\label{table:tabl2} The chemical shifts of energies of the $2p_{1/2}{\to}1s_{1/2}$ and $5d_{3/2}{\to}4p_{1/2}$ transitions and partial wave charges in the ionic and excited states of the isolated Pb atom.} 
\begin{ruledtabular}
\begin{tabular}{cccccccc}
        & & \multicolumn{2}{c} {Cations Pb$^{+n}$} &  \multicolumn{3}{c} {Excited states of Pb$^*$} & \\
         \cline{3 - 4}  \cline{5 - 7} \\
Transition & $q^W_{p1/2}$ \footnotemark[1]& $N^W_{6p1/2}$\footnotemark[2] & $\chi$\footnotemark[3], meV & $N^W_{6p1/2}$\footnotemark[2] & $N^W_{7p1/2}$\footnotemark[2] & $\chi'$\footnotemark[3], meV& $\delta$\footnotemark[4], \% \\
        \hline
        \multirow{3}{*}{$2p_{1/2}{\to}1s_{1/2}$ }  &0.0042 & 0.60 & -70   & 0.47   & 1.53 &  -72  & 2.8 \\
                                                   &0.0062 & 1.00 & -106  & 0.97   & 1.03 & -105  & 0.9 \\ 
                                                   &0.0097 & 1.80 & -151  & 1.77   & 0.23 & -150  & 0.7 \\
        \\
\multirow{3}{*}{$3p_{1/2}{\to}2s_{1/2}$}           &0.0042 & 0.60  & -147  & 0.47   & 1.53 &  -152  & 3.4 \\
                                                   &0.0062 & 1.00  & -221  & 0.97   & 1.03 &  -220 & 0.4 \\ 
                                                   &0.0097 & 1.80  & -313 & 1.77   & 0.23 &  -311 & 0.6 \\
\\
        \multirow{3}{*}{$5d_{3/2}{\to} 4p_{1/2}$ } &0.0042 & 0.60 & -422    & 0.47   & 1.53 &  -434  & 2.8 \\
                                                   &0.0062 & 1.00 & -633    & 0.97   & 1.03 &  -630  & 0.5 \\
                                                   &0.0097 & 1.80 & -897   & 1.77   & 0.23 &  -891 & 0.7 \\
\end{tabular}
\end{ruledtabular}
\footnotetext[1]{The $q^W_{p1/2}$ values are the partial wave charges of the electrons on the $6p_{1/2}$ and $7p_{1/2}$ states within the core region $r<R_c=0.5$.}
\footnotetext[2]{The $N^W_{nlj}$ values are the occupation numbers of the corrsponding one electron states.}
\footnotetext[3]{The $\chi$ and $\chi'$ values are the chemical shifts of the considered transition energies ( see \cite{Lomachuk:13} for details), with respect to the Pb$^{+2}$ isolated ion, in the ionic and excited states of the Pb atom correspondingly.}
\footnotetext[4]{The  $\delta = \frac{|\chi-\chi'|}{\chi}\cdot 100 \%$  value is the relative difference of the chemical shifts in ionic and excited states of the Pb atom.}
\footnotetext[0]{\\
The calculations were carried out with {\sc hfd} code \cite{hfd}.  The Pb shells from $1s$  to $5d$ were taken from Pb$^{+2}$ isolated ion calculations.}
        \end{minipage}
\end{table*}

It is shown in Ref.~\cite{Lomachuk:13} that the XES chemical shifts between two compounds can be calculated as the differences of mean values of an effective one-electron operator in these compounds. In \Table{table:tabl2} the evaluated chemical shifts of energies of the $2p_{1/2}{\to}1s_{1/2}$, $3p_{1/2}{\to}2s_{1/2}$, and $5d_{3/2}{\to}4p_{1/2}$ transitions with respect to Pb$^{2+}$ for various ionic and excited states of an unbound Pb atom are listed together with the values of partial wave charges $q^W_{p1/2}$ of $p_{1/2}$ spinors in the considered states. These states (electronic configurations) differ by only the occupation numbers of W shells, which are chosen in such a way that the partial wave charges
are the same in both ionic and excited states. Due to the W proportionality (\ref{eq:valorb}) the chemical shifts depend on the partial wave charges only. One can see from \Table{table:tabl2} that the chemical shift of the ionic and excited states with the same partial wave charges agree well to each other with the highest relative error of 10\% for the $5d_{3/2}{\to}4p_{1/2}$ transition. Note here that our formulation cannot be attributed to one of the four classes (charge models I--IV), discussed by Cramer in \cite{Cramer:04} since it is well defined theoretically  and reproduce well the experimental data.

\begin{table*}
        \begin{minipage}{\textwidth}
\caption{The partial wave charges and chemical shifts of the Pb $K-$lines with respect to the Pb$^{+2}$ ion
for PbH$_4$, PbF$_4$, PbO molecules and Pb$^{+4}$ ion.}
\label{table:tabl3}
\begin{ruledtabular}
\begin{tabular}{lccccccc}
        \multicolumn{8}{c}{Dirac-Fock calculations}\\
        \hline
        &Pb$^{+4}$\footnotemark[5] & PbF$_4$ \footnotemark[5] &Pb$^{+2}$\footnotemark[5] &  PbH$_4$ \footnotemark[5]&  PbO\footnotemark[5]& Pb$_2$\footnotemark[5]&Pb\footnotemark[5]\\
        \hline\\
        $q^{W}_{s1/2}$\footnotemark[1]& 0 & 0.0089  & 0.0233  &  0.0124&0.0182 &0.0191 &0.0193\\
        $q^{W}_{p1/2}$\footnotemark[1]& 0 & 0.0036  & 0.0000 &  0.0059 &0.0056 &0.0035 &0.0104\\
        $q^{W}_{p3/2}$\footnotemark[1]& 0 & 0.0037  & 0.0000 &  0.0057 &0.0021 &0.0027 &0.0000\\
\\
{\it V-conf.}\footnotemark[2] & - & $6s^{0.72}6p_{1/2}^{0.44}6p_{3/2}^{0.52}$ & $6s^2 $& $6s^{1.16}6p_{1/2}^{0.91}6p_{3/2}^{1.13}$ &  $6s^{1.76}6p_{1/2}^{0.90}6p_{3/2}^{0.42}$ & $6s^{1.81} 6p_{1/2}^{0.54} 6p_{3/2}^{0.62}$ & $6s^{2.00} 6p_{1/2}^{2.00}$ \\
\\

        $\chi_{2p1/2\to 1s1/2}$, meV\footnotemark[3]     & 330 &79  & 0 & -72 & -77 & --\footnotemark[8] & -175 \\
        $\chi_{2p3/2\to 1s1/2}$, meV\footnotemark[3]     & 391 &156 & 0 & -68 & -68 & -- & -156 \\
\\

$\chi_{2p1/2\to 1s1/2, rel}$, meV\footnotemark[4]&205&37 & 0 & -59& -82&-- &-150 \\
$\chi_{2p3/2\to 1s1/2, rel}$, meV\footnotemark[4]&222&45 & 0 & -55& -72&-- &-125 \\
\\
$\chi_{2p3/2\to 1s1/2, exp}$, meV\footnotemark[6]&   &   &   &    &     -$102 \pm 8$  & &\\
        \hline
        \multicolumn{8}{c}{Density functional theory calculations}\\
        \hline
        &Pb$^{+4}$ \footnotemark[7]& PbF$_4$ \footnotemark[7]  & Pb$^{+2}$ \footnotemark[7]& PbH$_4$ \footnotemark[7]&  PbO \footnotemark[7]  & Pb$_2$ \footnotemark[7]& Pb\footnotemark[7]\\
\hline\\
\\
$q^{W}_{s1/2}$ \footnotemark[1]&  0       & 0.0120    & 0.0250    & 0.0150  & 0.0198 & 0.0196 & 0.0207\\
$q^{W}_{p1/2}$ \footnotemark[1]&  0       & 0.0040    & 0.0000    & 0.0065  & 0.0053 & 0.0087 & 0.0109\\
$q^{W}_{p3/2}$ \footnotemark[1]&  0       & 0.0037    & 0.0000    & 0.0059  & 0.0027 & 0.0017 & 0.0000\\
\\
{\it V-conf.} \footnotemark[2] & -- & $ 6s_{1/2}^{1.07} 6p_{1/2}^{0.53} 6p_{3/2}^{0.57}$ & $6s_{1/2}^{2.00}$ & $6s_{1/2}^{1.53} 6p_{1/2}^{1.28} 6p_{3/2}^{1.69}$ & $6s_{1/2}^{2.00} 6p_{1/2}^{0.93} 6p_{3/2}^{0.61}$ & $6s_{1/2}^{2.00} 6p_{1/2}^{1.97} 6p_{3/2}^{0.59}$ & $6s_{1/2}^{2.00}  6p_{1/2}^{2.00}$\\
\\
$\chi_{2p1/2\to 1s1/2}$, meV\footnotemark[3]& 349 & 39       &  0        & -73       & -82     & -144\footnotemark[8]   &-187\\
$\chi_{2p3/2\to 1s1/2}$, meV\footnotemark[3]& 413 & 58       &  0        & -67       & -77     & -131   &-168\\
\\
$\chi_{2p1/2\to 1s1/2, rel}$, meV\footnotemark[4]& 205&6&0&-94&-96&-140&-151\\
$\chi_{2p3/2\to 1s1/2, rel}$, meV\footnotemark[4]& 223&0&0&-90&-90&-122&-125\\
\end{tabular}
\end{ruledtabular}
\footnotetext[1]{ The $q^{W}_{lj}$ values are the partial wave charges values for the shells starting from $6s$ in compounds within the sphere of $R_c = 0.5$~a.e.\ radius centered on the Pb atom.}
\footnotetext[2]{ The occupation numbers of the valence states of the Pb atom obtained from partial wave charges values (see \Eq{confs_iter} and text below.}
\footnotetext[3]{ The $\chi_{fi}$ values are the chemical shifts of XES lines corresponding to the transitions between the $F$ and $I$ shells of the Pb atom in the given compound with respect to the Pb$^{+2}$ ion. These values are computated by the method described in Ref.~\cite{Lomachuk:13}.}
\footnotetext[4]{ The $\chi_{fi, rel}$ values are the values of the XES chemical shifts of the transitions between the $F$ and $I$ shells of the Pb atom in the given compound with respect to the Pb$^{+2}$ ion, obtained within the relativistic average configuration calculations of the isolated ions, carried out with help of the {\sc hfd} code\cite{hfd}. The isolated ion electronic configurations are correspond to the configurations listed in these table. This way of the computation of the chemical shifts allows one to take into the account the core shells changing from one compound to another.} 
\footnotetext[5]{ The electronic structure of the Pb$_2$, PbH$_4$, and PbF$_4$ molecules, and Pb$^{+2}$ isolated ion were calculated by the {\sc dirac} code. These calculations were carried out in the Dirac--Fock approximation with using of semilocal 22 electron relativistic pseudo potential\cite{Petrov:05a}, the Pb spinors, that belonging to the shells from $5s$ to $5d$  are frozen and taken from Pb$^{+2}$ computation.}
\footnotetext[6]{ The value obtained from experimental data for the chemical shifts of the $K\alpha_1$  lead XES lines in the PbO crystal with respect to the crystalline metallic lead \cite{Egorov:92}. The presented experimental value is obtained as difference between the experimental chemical shift value and chemical shift of the Pb$_2$ molecule presented in the $\chi_{2p3/2\to 1s1/2, rel}$ row.} 
\footnotetext[7]{ The results of calculations are obtained with the {\sc dirac} code in the DFT framework, the used functional is the PBE0 \cite{pbe0}. The Pb spinors belonging to the shells from $5s$ to $5d$ were taken from Pb$^{+2}$ calculations. The semilocal 22-electron relativistic effective core potential \cite{Petrov:05a} was used.}
%
\footnotetext[8]{The electronic correlations taken into account at the DFT level lead to the Pb$_2$ ground state configuration different from that obtained at the Dirac--Fock level. This can be seen from comparison of the corresponding occupation numbers listed in the ``{\it V-conf.}'' rows and the partial wave charges $q^W_{p1/2}$. The ground state configuration of the Pb$_2$ valence electrons at the DFT level is $\pi^2\sigma^2$, while the configuration of valence electrons at the Dirac--Fock level is $\pi^2\pi^2$. As a result, the XES chemical shifts obtained at the DFT level are almost three time more than those calculated at the Dirac--Fock level. The comparison of the Dirac--Fock and DFT chemical shifts is uninformative here and the Dirac--Fock chemical shifts are not listed in the table.} 
\end{minipage}
\end{table*}

In \Table{table:tabl3} the partial wave charges $q^{W}_{lj}$ and effective occupation numbers of the valence Pb (sub)shells in the PbH$_4$, PbF$_4$, and Pb$_2$ compounds and Pb$^{+2}$, Pb$^{+4}$ ions are listed together with the chemical shifts of the x-ray $K$-line in Pb with respect to the Pb$^{+2}$ ion. The partial wave charges and XES chemical shifts values are obtained with using the two-step restoration codes developed in Refs.\ \cite{Skripnikov:13b,Skripnikov:13c} after pseudopotential calculations carried out with the {\sc dirac} code \cite{dirac13}.  Due to the stability of the computation procedure issues, calculations of the electronic structure of all listed compounds were performed with the core Pb shells frozen up to $5d$. The states belonging to these shells were taken from the Pb$^{+2}$ computation. The effective occupation numbers of the valence shells, $N^W_{n_vlj}$, were determined from the following equations:
\begin{equation}
N^W_{n_vlj}\int\limits_{r<R_c} (\left|\phi^L_{n_vlj}(r)\right|^2 + \left|\phi^S_{n_vlj}(r)\right|^2)r^2\,dr = q^W_{lj},
\label{confs_iter}
\end{equation}
where $\phi^L_{n_vlj}(r)$ and $\phi^S_{n_vlj}(r)$ are the radial functions of the large and small components of the corresponding valence state obtained from the relativistic average configuration computation of the isolated ion with frozen core states; the occupation numbers of the valence Pb shells in this computation are equal to the $N^W_{n_vlj}$ values. Thus \Eq{confs_iter} is a nonlinear self-consistent equation and must be solved iteratively. Corresponding calculations were carried out with the {\sc hfd} code \cite{hfd}.

It is possible to carry out all-electron calculations of an isolated ion with the given occupation numbers, $N^W_{n_vlj}$, and to take partially account of relaxation of the core states frozen earlier.  The chemical shifts obtained from these calculations are also listed in \Table{table:tabl3}. The differences of these values and the chemical shifts obtained in the frozen core calculations are 10\%--20\% by the order of magnitude for the neutral Pb atom and molecules PbH$_4$, PbO, Pb$_2$, which are weaker bound compared to PbF$_4$. For the Pb$^{+4}$ ions and PbF$_4$ the relaxed chemical shifts are about two to three times lower than the corresponding values obtained in the frozen core calculations. 

The experimental datum for the chemical shift of the XES $2p_{3/2}\to 1s_{1/2}$ line of lead in crystalline PbO with respect to metallic Pb is listed in Ref.~\cite{Egorov:92} and equals to $54\pm8$ meV. One can estimate the corresponding value from the performed calculations as difference between chemical shifts of PbO and Pb$_2$. For the results obtained in the Dirac--Fock approximation this estimate gives 9 meV without taking into the account the Pb core relaxation in the compounds and it is 15 meV for the relaxed core case. 

It is possible to take account of the effects of electronic correlation within DFT. From the results of calculations with using the PBE0 functional \cite{pbe0} listed in \Table{table:tabl3} we conclude that the electronic correlation effects are important for the Pb$_2$ molecule, since the XES chemical shifts differ by the factor of two for Dirac--Fock and DFT calculations. The chemical shift of the lead XES $2p_{3/2}\to 1s_{1/2}$ line obtained at the DFT level is 54 meV when the core relaxation is not considered and is 32 meV when the relaxation is taken into account. The obtained values are much closer to the experimental datum than those obtained at the Dirac--Fock level.



\section{Conclusions}

Utilizing the property of proportionality of valence and low-lying virtual spinors within an atomic core region with radius $r{<}R_c$, the notions of ``W reduced density matrices'', ${\bm\rho}^W$, and ``partial wave charges'', $q^{W}_{lj}$, for valence electrons in the core region are introduced. Such properties as hyperfine structure constants
\comtav{for TableI+: <<<}
    (\Table{table:tabl1}),
T- and P-violation effects (\cite{Titov:06amin}), XES chemical shifts (\Table{table:tabl3}),
\comtav{>>> End}
which are mainly sensitive to a variation of electronic densities in an atomic core region (or even on a nucleus), with a good accuracy depend on ${\bm\rho}^W$ only. For specific AiC properties or effective Hamiltonian parameters the more particular blocks of ${\bm\rho}^W$ like diagonal terms $q^{W}_{lj}$ (for XES chemical shifts), off-diagonal $s{-}p$, $p{-}d$, etc.
submatrices (for evaluation of \Eeff\ and some other P- and T,P-odd Hamiltonian parameters with the $sp$-, $pd$-, etc.\ hybridized unpaired electrons \cite{Kozlov:87, Dmitriev:92})  are sufficient to know. Thus, the $W$ reduced DM allows one to characterize the effective state of an atom in a chemical substance by an appropriate manner.

The features of the AiC approach are summarized as follows.\\
{\bf (i)} The $W$ reduced DM and, correspondingly, the AiC characteristics calculated on an atom are independent of the origin of one-electron basis set
used (whether it is a one-center, MO~LCAO, analytic, or numerical one) in the limit of its completeness in contrast to the cases of Mulliken and L\"owdin population analyses; the one-center AiC basis functions are independent on the valence structure of a chemical substance studied in contrast
to those in the NAO approach.\\
%
{\bf (ii)} it describes well (generally, in the range of accuracy of 10-30\%) the multitude of AiC characteristics.\\
%
{\bf (iii)} Due to the property of $W$ proportionality, the core radius $R_c$ is not a very critical parameter to be fixed as exactly as possible for an AiC characteristic but the accuracy of calculation is higher for those characteristics for which it can be chosen smaller since the $W$ reduced DM match better the original DM for smaller $R_c$%
%
\footnote{The $W$ proportionality can be interpreted as some kind of ``asymptotic {\it unfreedom}'' 
 of valence and low-lying virtual spinors in an atomic core of a chemical substance 
in contrast to a widely known property of ``asymptotic freedom'' in theory of strong interactions.};
the situation is similar to that in theory of ``transferable'' shape-consistent relativistic pseudopotentials with the $R_c$ treated as a matching radius (see Refs.~\cite{Goedecker:92a, Titov:99} and references).
\\
{\bf (iv)} The method allows one to give a correct quantum mechanical interpretation even for such difficult cases as XES chemical shifts and provide an unambiguous analysis of atomic (one-center) spinor contributions to AiC characteristics in complicated electronic structures (see discussion in \cite{Kudashov:13} concerning RaO);\\
{\bf (v)} it can provide a theoretical background for semiempirical models to evaluate (estimate) the AiC characteristics which are not known or cannot be calculated using available experimental data for corresponding properties \cite{Kozlov:95}.\\

Note as well that the approach can be easily implemented in codes when  the relativistic pseudopotential theory \cite{Titov:99, Petrov:04b, Mosyagin:06amin} and one-center recovery (restoration) procedures \cite{Titov:05a, Titov:06amin} can be utilized for calculation of the AiC properties and parameters in heavy-element compounds.
However, within the AiC approach, the concept of a charge on an atom in a chemical substance becomes meaningless \cite{Cramer:04}. 
%

\section*{Acknowledgement}
   This work is supported by the Russian Science Foundation grant No.~14-31-00022.
We are grateful to A.Zaitsevskii for fruitful discussions.


\bibliographystyle{./bib/apsrev}
\bibliography{./bib/JournAbbr,./bib/QCPNPI,./bib/TitovLib,./bib/Lomachuk,./bib/SkripnikovLib}


\end{document}